# تحسين بروتوكول LEACH ليتكيّف مع بعد الحساسات عن المحطة الأساسية

م.أحمد الهلال[1] , د. صلاح الدوه جي[2]

[1] Faculty of Information Technology Engineering, Damascus University, Syria

[1] طالب ماجستير – قسم النظم والشبكات الحاسوبية، كلية الهندسة المعلوماتية, جامعة دمشق, سوريا

[2] Associated professor, Faculty of Information Technology Engineering, Damascus University, Syria

[2] أستاذ مساعد, كلية الهندسة المعلوماتية, جامعة دمشق, سوريا

**الملخص:**

تحتاج بعض التطبيقات إلى نشر حساسات في حقل العمل لتقوم هذه الحساسات بتحسُّس البيئة وتوجيه المعطيات المحسوسة إلى المحطة الأساسية حيث تتم معالجتها, وهذه الحساسات تعتمد على بطاريات غير قابلة للتبديل أو إعادة الشحن, ولذلك فإنَّ خوارزميات التوجيه المُقترحة للشبكات من هذا النوع من العقد (الحساسات) يجب أن تكون فعّالة للطاقة. أحد البروتوكولات المقترحة لهذه الشبكات هو بروتوكول التوجيه الهرمي **LEACH** الذي يُساعد على توفير الطاقة ويتكيّف مع الطاقة المنخفضة في هذه الحساسات.

سنقوم في هذا البحث بدراسة شبكات الحساسات المتجانسة و تحسين بروتوكول **LEACH** (الذي يعمل في الأساس بكفاءة وينتج أداء عالي في هذا النوع من الشبكات) ليتكيّف مع البعد عن المحطة الأساسية, وبالتالي توفير أكثر في الطاقة من أجل أبعاد للمحطة الأساسية عن حقل العمل.

يعتمد بروتوكول **LEACH** المُحسَّن على نموذج رياضي لحساب معدّل طاقة الشبكة بشكل تقديري في كل جولة من جولات العمل وبالتالي الإستفادة من الطاقة المتبقية في العقد في توزيع دور رأس العنقود **Cluster Head** على باقي العقد, ويعتمد أيضاً على نموذج رياضي لحساب بعد المحطة الأساسية عن حقل العمل, بينما لا يأخذ بروتوكول **LEACH** أي اعتبارات للطاقة المتبقية في العقد.

ستوضِّح نتائج المحاكاة أنَّ بروتوكول **LEACH** المُحسَّن يعطي أداء أفضل من **LEACH** في الحالات التي تكون فيها المحطة الأساسية ضمن حقل العمل والأداء يتمثّل في زيادة استقرار ووثوقية الشبكة, ويتماثل مع أداء **LEACH** في الحالات التي تكون فيها المحطة الأساسية خارج حقل العمل.

**الكلمات المفتاحية:** شبكات الحساسات اللاسلكية, البيئة المتجانسة, فعاليّة الطاقة, بروتوكولات التوجيه الهرميّة.

# Base-Station Distance Adaptive LEACH

Ahmad Alhilal[1] and Salah Dowaji[2]

[1] Faculty of Information Technology Engineering, Damascus University, Syria

[2] Associated professor, Faculty of Information Technology Engineering, Damascus University, Syria

**Abstract:**
*For some applications, we need to deploy a network of sensors in working field to sense the environment and send collected data to a base-station for processing; these sensors depend on non rechargeable batteries, so the routing protocols for a such network of sensors need to be efficient. LEACH is one of these protocols which is a hierarchical routing protocols and helps in saving energy in wireless sensor networks.*

*Enhanced LEACH depends on mathematical model to calculate an estimated average energy in each round consequently make using on node's remaining energy to ensure rotating cluster head role across all nodes, It also depends on mathematical model to calculate base station distance from work field whereas LEACH doesn't take into its account any consideration for remaining energy of node.*

*In this paper, we enhance LEACH (work efficiency in homogeneous networks) to adapt with base-station distance, thus more energy saving for certain distances from base-station. The obtained simulation results show that enhanced LEACH saves energy better than LEACH and increase network stability and reliability when base-station is inside working field and consume same energy as LEACH when base-station is outside work field.*

# 1. المقدمة:

تُعتبر شبكات الحساسات من التقنيات الحديثة في عالم التكنولوجيا، حيث تتيح هذه الشبكات طيف واسع من التطبيقات الجديدة المُتمثِّلة في ربط منظومات التحكم والمراقبة بالعالم الحقيقي عن طريق نشر حساسات قريبة من الظاهرة أو الحدث قيد الدراسة، وتقوم هذه الحساسات بجمع المعطيات و إرسالها إلى مركز اتخاذ قرار(المحطة الأساسية) والذي يقوم بدوره بتحليلها ومن ثم اتخاذ القرار المناسب.

وتُعرَّف شبكات الحساسات اللاسلكية على أنَّها مجموعة من العقد الحساسة التي تتصل مع بعضها لاسلكياً دون الحاجة إلى بنية تحتية وتعتمد العقد الحساسة على بطاريات لتزويدها بالطاقة, ومن أهم وظائفها استشعار الوسط المحيط ونقل المعطيات المحسوسة عن طريق عقد تمرير أو بدونها إلى العقدة الوجهة (المحطة الأساسية), كما هو وارد في المراجع [2-4].

تتميز شبكات الحساسات اللاسلكية بمحدودية موارد العقد المكونة لها من حيث مدى الإرسال وعرض الحزمة المتاحة وقدرة هذه العقد على المعالجة والتخزين، بالإضافة إلى محدودية موارد الطاقة التي تغذِّي هذه العقد والتي تتمثَّل ببطارية غير قابلة لإعادة الشحن أو التبديل في معظم الحالات.

## 1.1. تطبيقات هذه الشبكات :

➢ مراقبة البيئة : رطوبة التربة, درجات الحرارة...
➢ تطبيقات عسكرية: الكشف عن الهجمات البيولوجية والكيميائية ومراقبة ساحة المعركة.
➢ تطبيقات صناعية : مراقبة خطوط الانتاج وضبطها في الاماكن التي يصعب وصول الانسان إليها.
➢ تطبيقات طبية: رصد البيانات الفيزيولوجية عن بعد.
➢ تطبيقات البيت الذكي : بيئة ذكية.

## 2.1. أنواع بروتوكولات التوجيه:

قدَّمت الأبحاث [3,4] تصنيفاً لبروتوكولات التوجيه في شبكات الحساسات اللاسلكية وفق الأنواع التالية :

1.2.1. التوجيه المسطَّح (Flat routing):
تعتمد على استعلامات تُنشأ من قبل المحطة الأساسية ومن ثم تُنشر عبر الشبكة.

2.2.1. التوجيه الهرمي (Hierarchical routing):

تقوم هذه البروتوكولات بتقسيم الشبكة إلى مجموعة من العناقيد، وتتعامل مع نوعين من العقد، النوع الأول العقد العادية التي تقوم بمهام استشعار الوسط، والنوع الثاني عقد تعمل رؤوساً للعناقيد وتقوم هذه الأخيرة بتجميع البيانات الواردة من العقد وتجميعها و إرسالها إلى المحطة الأساسية(BS) حيث تتم معالجتها و الإستفادة منها في التطبيق الذي وُظِّفت من أجله كما في الشكل(1).

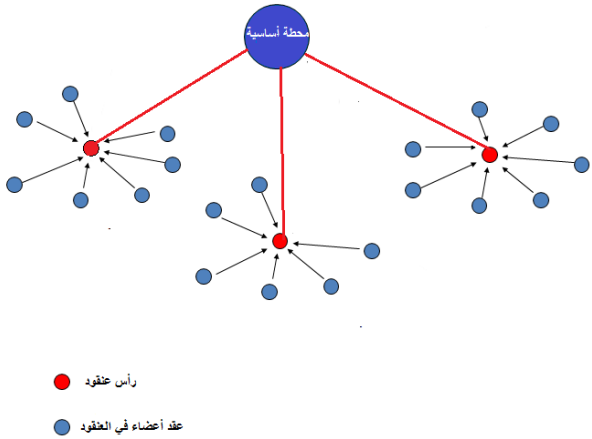

الشكل ( 1 ) تقسيم شبكة الحساسات في بروتوكولات التوجيه الهرمية

هناك عدة بروتوكولات مُستخدمة في هذا النطاق نذكر منها:

➢ Low Energy Adaptive Clustering Hierarchy (**LEACH**): خوارزمية توجيه هرمية تعتمد على العناقيد للتكيُّف مع الطاقة المنخفضة.

➢ Power–Efficient Gathering in Sensor Information Systems (**PEGASIS**):
فعاليّة طاقة جمع المعلومات في أنظمة الإستشعار[8].

➢ Threshold sensitive Energy Efficient sensor Network protocol (**TEEN**):
خوارزمية توجيه هرمية فعّالة بالنسبة للطاقة وحسّاسة للعتبة.

3.2.1. التوجيه المعتمد على موقع العقد (Location based routing): تعتمد على معلومات مواقع العقد بهدف نقل المعطيات عبر الشبكة باتجاه العقدة الوجهة.

4.2.1. التوجيه المعتمد على جودة الخدمة (QOS based routing): تأخذ قضايا جودة الخدمة بعين الإعتبار نذكر مثلاً زمن التأخير طرف-طرف عند بناء المسارات في شبكة الحساسات اللاسلكية.

## 1.3. النموذج الراديوي لتحليل الطاقة:

سنعتبر نموذج راديوي بسيط لتوصيف الطاقة المبدّدة من عتاديات الأشعة الراديوية, فالحساس المُرسل يبدد طاقة لتشغيل الالكترونيات الراديوية وطاقة لتضخيم الطاقة لأيصال البيانات إلى مسافات أطول, بينما المستقبل يبدد طاقة لتشغيل الإلكترونيات الراديوية كما في الشكل (2) .

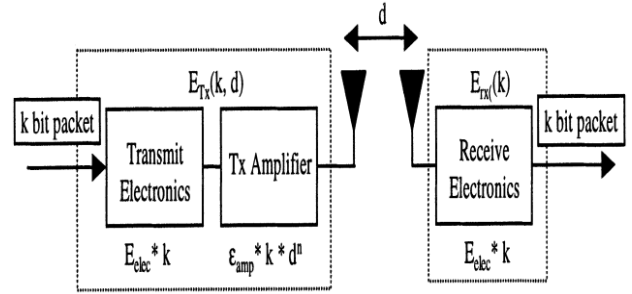

الشكل ( 2 ) الإرسال والإستقبال في الحساسات

تستخدم بروتوكولات التوجيه في هذه الشبكات نوعين من نماذج القنوات كطريقة لفقدان الطاقة :

- فقدان الطاقة في حال فضاء حر وتحتاج طاقة تضخيم $\varepsilon_{fs}d^2$ حيث $\varepsilon_{fs}$ هي طاقة التضخيم في حال فضاء حر و d هي البعد بين المرسل والمستقبل.

- فقدان الطاقة في حال تخامد متعدد المسارات $\varepsilon_{mp}d^4$ حيث $\varepsilon_{mp}$ هي طاقة التضخيم في حال استخدام نموذج التخامد متعدد المسارات.

يمكن استخدام متحكم طاقة لإختيار أحد نماذج القنوات السابقة وذلك بإعداد مناسب لمضخم الطاقة كالتالي:

- إذا كان البعد بين المرسل والمستقبل أقل من عتبة $d_0$ عندها يُستخدم نموذج الفضاء الحر.

- وإلا يُستخدم نموذج التخامد متعدد المستويات.

تستهلك الأشعة الراديوية من أجل رسالة مؤلفة من k بت والبعد d مايلي :

**طاقة الإرسال :**

$$E_{TX}(k,d) = kE_{elec} + k\varepsilon_{fs}d^2 \quad (1)$$

وهو الشكل المبسَّط لطاقة الإرسال على مسافات قريبة , أما الشكل العام فهو :

$$E_{TX}(k,d) = \begin{cases} kE_{elec} + k\varepsilon_{fs}d^2 , d < 0 \\ kE_{elec} + k\varepsilon_{mp}d^4 , d \geq 0 \end{cases} \quad (2)$$

حيث k عدد بتات الرسالة , d مسافة الإرسال , $E_{elec}$ الطاقة الإلكترونية, $E_{\varepsilon mp}$ طاقة التضخيم.

**طاقة الإستقبال:**

تستهلك عتاديات الأشعة الراديوية طاقة لإستقبال الرسالة كالتالي:

$$E_{TX}(k,d) = kE_{elec} \quad (3)$$

حيث $E_{elec}$ هي طاقة الإلكترونيات وتعتمد على عوامل مثل الترميز الرقمي , التعديل , الفلترة و توسيع الإشارة, بينما يعتمد مضخم الطاقة $\varepsilon_{fs}d^2$ أو $\varepsilon_{mp}d^4$ على المسافة إلى المرسل ومعدل خطأ ترميز مقبول.

**الطاقة المستهلكة في رأس العنقود:**

يستهلك رأس العنقود طاقة لإستقبال الإشارات من العقد , وطاقة لتجميع الإشارات وطاقة لنقل الإشارة المجمّعة إلى المحطة الأساسية.

نعتبر مسبقاً أنّ الطاقة المبدّدة تتبع نموذج التخامد متعدد المسارات $d^4$ بما أن المحطة الأساسية بعيدة عن العقد, وبناءاً على ذلك فإنّ الطاقة المبددة في رأس العنقود لإرسال رسالة تُحسب كالتالي :

$$E_{CH} = kE_{elec}\left(\frac{N}{k} - 1\right) + kE_{DA}\frac{N}{K} + k\varepsilon_{mp}d_{toBS}^4 \quad (4)$$

حيث $d_{toBS}$ هو البعد بين رأس العنقود والمحطة الأساسية و $E_{DA}$ هي طاقة تجميع المعطيات و N هو عدد العقد.

**الطاقة المستهلكة في العقد غير الرأسية:**

تحتاج كل عقدة غير رأسية إلى نقل معطياتها إلى رأس العنقود فقط مرة من أجل كل إطار وبما أنّ البعد بين العقدة و رأس العنقود صغير وبالتالي فالطاقة المبدّدة تستخدم نموذج فقدان طاقة في فضاء حر , لذلك فالطاقة المستخدمة في هذه العقد تُحسب كالتالي:

$$E_{non-CH} = kE_{elec} + k\varepsilon_{fs}d_{toCH}^2 \quad (5)$$

حيث $d_{toCH}$ هو البعد بين العقدة ورأس العنقود.

سندرس في هذا البحث بروتوكول التوجيه الهرمي LEACH ونقوم بالتحسين عليه إنطلاقاً من مساوئه.

5. محتوى البحث:

قدّم البحث [1] الذي اقترحه الباحثان A. Alhilal, S. Dowaji تقييماً لأداء بروتوكولات التوجيه الهرمية حسب مستوى التجانس والذي استنتج الباحثان من خلاله بأنَّ بروتوكول التوجيه الهرمي LEACH هو البروتوكول الأكثر فعاليّة من حيث أدائه في شبكة الحساسات اللاسلكية المتجانسة, لذلك سينطلق هذا البحث من هذا البروتوكول ويحاول التحسين عليه بحيث نحصل على أداء أعلى للشبكة (دورة حياة, فترة استقرار, فترة عدم استقرار), وسنستخدم LEACH للمقارنة مع البروتوكول الذي سينتجه البحث.

### 2.1. بروتوكول **LEACH** :

كما هو وارد في المرجع [5], اقترح الباحثان W. R. Heinzelman, A. P. Chandrakasan بروتوكول LEACH والذي يعتمد في انتخابه لرؤوس العناقيد على عتبة تُحسب وفق العلاقة التالية:

$$T(s) = \begin{cases} \frac{Popt}{1-Popt(r \bmod \frac{1}{Popt})} & if\ s \in G \\ 0 & otherwise \end{cases} \quad (6)$$

نعرّف أولاً عمليات الشبكة على أنّها عمليات انتخاب رؤوس العناقيد وتحسّس البيانات وإرسالها وإستقبالها ويتم تقسيم عمليات الشبكة إلى جولات (**rounds**) والجولة تعني فترة زمنية معينة يتم في بدايتها عمليات إنتخاب جديدة لرؤوس العناقيد.

في المعادلة (6) تمثّل **r** رقم الجولة الحالية, Popt هي الإحتمالية الأمثلية التي يتم انتخاب رؤوس العناقيد وفقها وهي قيمة تجريبية تمّ تحديدها في برمترات المحاكاة على أنّها 0.05, **G** هي مجموعة العقد التي لم تُنتخب رؤوساً للعناقيد في آخر epoch(فترة مكوّنة من عدة جولات) واحتمالية هذه العقد تزداد بعد كل جولة في نفس epoch .

يقوم بروتوكول LEACH باختيار رقم عشوائي ضمن المجال [0,1], فإذا كان الرقم أصغر من العتبة عندها يتم اختيار العقدة كرأس للعنقود.

نلاحظ بإنّ بروتوكول LEACH يعتمد على قيمة عشوائية ومجموعة العقد **G** لإنتخاب رأس العنقود وهو الأمر الذي لا يضمن الإنتخاب بطريقة عادلة (الإنتخاب حسب الطاقة المتبقية الأكبر) وبالتالي عدم استقرار في الشبكة حيث تتبدد طاقة عقد مبكراً وتستمر طاقة عقد أخرى لفترات أطول.

### 2.2. بروتوكول **LEACH** المتكيّف للبعد عن المحطة الأساسية:

تُحسب العتبة في هذا البروتوكول على غرار تلك التي في بروتوكول LEACH وفق العلاقة (6), ولكنه لايعتمد فقط على هذه العتبة لإنتخاب رأس العنقود فهو يعتمد أيضاً على معدل طاقة الشبكة وبالتالي فإنه يعتمد في الإنتخاب على مايلي:

1. يقوم أيضاً باختيار رقم عشوائي ضمن المجال [0,1], فإذا كان الرقم أصغر من العتبة عندها يختبر الشرط الثاني.

2. طاقة العقدة أكبر من معدل طاقة الشبكة المقدّرة.

يمكننا نظرياً حساب الطاقة الوسطية للشبكة (معدل طاقة الشبكة) في الجولة رقم (**i**) كالتالي :

حساب الطاقة الوسطية في الجولة رقم (**i**) = مجموع طاقات العقد في الجولة رقم (**i**) مقسوماً على عدد العقد الحية وهي تُعبِّر عن الطاقة الوسطية الحقيقة للشبكة. فعلياً لا يمكننا حساب هذه القيمة في كل عقدة لأنّها تتطلب معرفة طاقات باقي العقد في الشبكة وطبعاً هذا صعب ويولّد عبء على الشبكة واستهلاك أكبر في طاقة العقد ولتجنب ذلك سنلجأ إلى تقدير معدل طاقة الشبكة كما اقترحها الباحثون Li Qing, Qingxin Zhu, Mingwen Wang في المرجع [6].

#### 2.2.1. تقدير معدل طاقة الشبكة:

تقوم كل عقدة بحساب طاقة الشبكة الكلية وفق العلاقة التالية:

$$E_{total} = NE_0 \quad (7)$$

تنطلق كل عقدة في حساب الطاقة الكلية من كون الشبكة متجانسة وبالتالي كل العقد تمتلك نفس الطاقة لإبتدائية $E_0$ , ثم تقوم كل عقدة بتقدير معدل طاقة الشبكة دون الحاجة إلى معرفة طاقة بقية العقد وذلك حسب العلاقة التالية:

$$\bar{E}(r) = \frac{1}{N} E_{total} \left(1 - \frac{r}{R}\right) \quad (8)$$

حيث R هو عدد الدورات الأعظمي الذي يمكن أن تستغرقها عمليات الشبكة ويتم حسابها وفق العلاقة التالية:

$$R = \frac{E_{total}}{E_{round}} \quad (9)$$

تُحسب الطاقة المستهلكة في كل جولة وفق العلاقة التالية:

$$E_{round} = L(2NE_{elec} + NE_{DA} + k\varepsilon_{mp}d_{toBS}^4 + N\varepsilon_{fs}d_{toCH}^2) \quad (10)$$

حيث $E_{elec}$ طاقة الإرسال, $E_{DA}$ طاقة التجميع, $d_{toBS}$ البعد بين رأس العنقود والمحطة الأساسية, $d_{toCH}$ البعد بين رأس العنقود وأعضائه.

### 2.2.2. حساب بعد المحطة الأساسية عن حقل العمل:

نُعرِّف أولاً عتبة البعد على أنّها العتبة التي نقرر من خلالها فيما إذا كانت المحطة الأساسية خارج حقل العمل أو ضمنه وتُحسب وفق العلاقة التالية: $d_{th}$=

$$d_{th} = \sqrt{\frac{\varepsilon_{fs}}{\varepsilon_{mp}}} \quad (11)$$

حيث $E_{fs}$ هي الطاقة اللازمة لتضخيم طاقة الإرسال في حال استخدام النموذج الراديوي في فضاء حر و $E_{mp}$ هي الطاقة اللازمة لتضخيم طاقة الإرسال في حال استخدام النموذج الراديوي في فضاء متعدد المسارات.

يُحسب بعد العقدة عن المحطة الأساسية وفق المعادلة التالية:

$$BS_{distance} = \sqrt{(S(i).x - Sink.x)^2 + (S(i).y - Sink.y)^2} \quad (12)$$

نقارن بعد العقدة عن المحطة الأساسية $BS_{distance}$ مع عتبة البعد $d_{th}$ لمعرفة الطريقة التي سوف تستخدم في إنتخاب رؤوس العناقيد، وبالتالي يمكننا تحديد فيما إذا كانت المحطة الأساسية داخل أو خارج حقل العمل, فإذا كانت خارج حقل العمل عندها يتم تطبيق منطق خوارزمية LEACH التقليدية للإنتخاب وإلا سيتم تطبيق منطق خوارزمية LEACH المُحسّنة للقيام بالإنتخاب.

فتكون الخوارزمية الناتجة (pseudo Code) وفق الشكل التالي:

```
For each round{
Calculate the energy spent in each round Er.
Calculate Estimated total energy Et=Et–Er.
Calculate Estimated average energy Ea.
    • Cluster Head Election
For each node{
If (BS inside work field){   // BSdistance<=dth
    If (Ea>٠ && (s(i).G<=٠))
    Select random number rand
        If(rand<T(s) && node's energy > average energy)
            Elect node as Cluster Head
    }
    Else { //BS outside work field
    If (Ea>٠ && (s(i).G<=٠)){
        Select random number rand
        If(rand<T(s)){
        Elect node as Cluster Head
            }
        }
    }
} //end for each node
```

### 3.2.2. المخطط التدفقي لخوارزمية LEACH المُحسّنة:

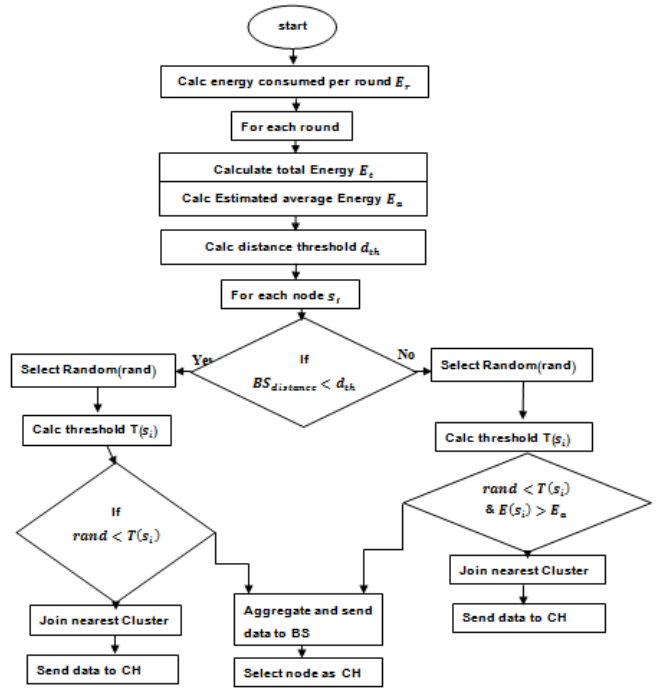

الشكل ( 3 ) مخطط تدفقي لخوارزمية LEACH المُحسّنة

تحسين بروتوكول LEACH ليتكيّف مع بعد الحساسات عن المحطة الأساسية

## 4. مقارنة بين بروتوكول LEACH و بروتوكول LEACH المُحسّن:

سندرس حالتين من موقع المحطة الأساسية هما:

### 1.4. الحالة الأولى: المحطة الأساسية ضمن حقل العمل

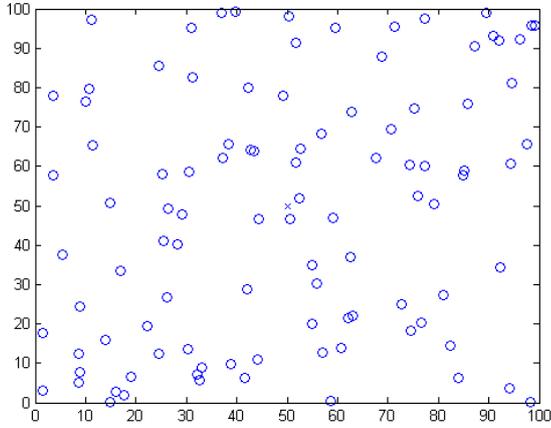

الشكل (5) نشر العقد وتموضع BS في وسط حقل العمل

تتوضَّع المحطة الأساسية في الشكل (5) في منتصف الحقل وذلك في النقطة (50,50).

سنقارن بين البروتوكولين حسب مقاييس الأداء التي ذكرها الباحثون G. Smaragdakis, I. Matta, A. Bestavros في المرجع [7].

### 4.1.1. دورة حياة الشبكة:

هي الوقت الفاصل بين بدء عمليات الشبكة وحتى موت آخر عقدة حية.

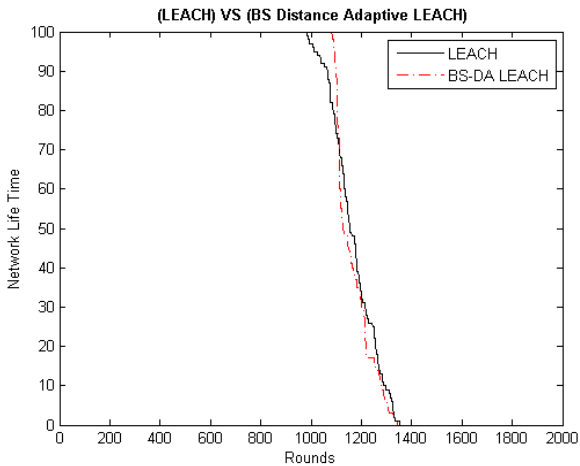

الشكل (6) مقارنة دورة حياة الشبكة بين LEACH و LEACH المحسّن

نلاحظ من الشكل (6) أنَّ دورة حياة الشبكة في LEACH تتماثل مع تلك التي في LEACH المُحسّن وهذا المقياس ليس

```
• Joining clusters
  If(node is normal){ //not CH
    Join nearest cluster
  }

• CH assignment
  If(node is CH){
    Aggregate data then send it to BS
  }

• Normal node assignment
  If(node is normal){ //not CH
    Send data to CH
  }
} //end rounds
```

الشكل (4) pseudo Code لخوارزمية LEACH المُحسّنة

## 3. بارامترات المحاكاة:

سوف نستخدم برنامج Matlab كوسيلة للمحاكاة والمقارنة بين أدائي بروتوكول LEACH و بروتوكول LEACH المحسّن وسنتقيد بوسطاء المحاكاة المستخدمة في LEACH والواردة في الجدول (1):

| البارامتر | معناه | القيمة |
|---|---|---|
| $E_{elect}$ | طاقة الكترونيات الإرسال والإستقبال | 5 nJ/bit |
| $\varepsilon_{mp}$ | طاقة نموذج تعدد المسارات | 10 pJ/bit/m² |
| $\varepsilon_{fs}$ | طاقة نموذج الفضاء الخالي | 0.0013 pJ/bit/m⁴ |
| $E_o$ | الطاقة الابتدائية | 0.5 J |
| $E_{DA}$ | طاقة التضخيم | 5 nJ/bit/msg |
| $d_o$ | عتبة البعد | 70 m |
| $P_{opt}$ | الاحتمال الأمثلي | 0.05 |
| K | حجم الرسالة | 4000 bits |
| n | عدد العقد | 100 |

الجدول (1) برمترات المحاكاة والمقارنة

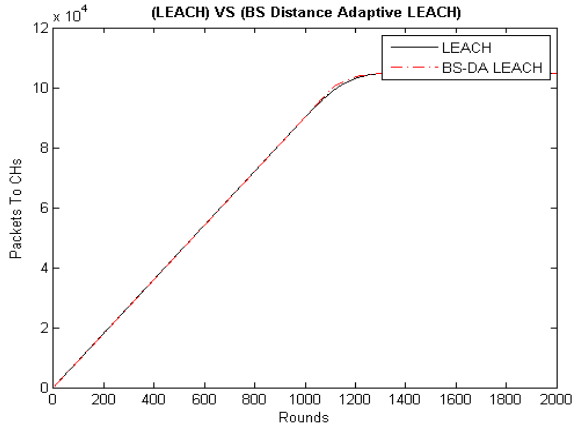

الشكل (8) عدد الطرود المُرسلة إلى رؤوس العناقيد

نلاحظ من الشكل (8) أنّ إنتاجية LEACH المُحسّن من حيث عدد الطرود المُرسلة إلى رؤوس العناقيد تتطابق مع انتاجيّة LEACH.

### 4.1.5. عدد رؤوس العناقيد في كل جولة:

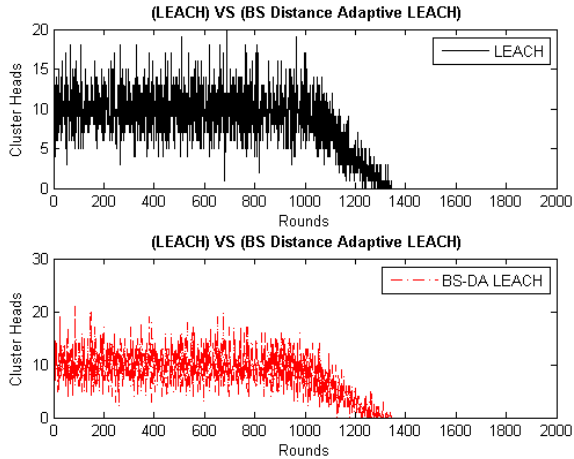

الشكل (9) عدد رؤوس العناقيد بكل جولة

نلاحظ تطابق عدد رؤوس العناقيد في كل جولة في كل من LEACH وLEACH المحسّن وذلك حسب الشكل (9).

### 4.2. الحالة الثانية : المحطة الأساسية خارج حقل العمل

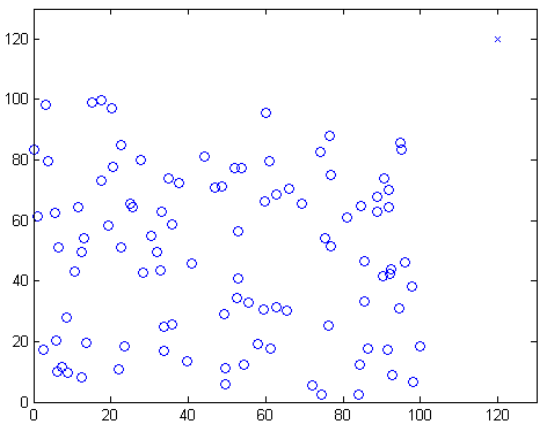

الشكل (10) نشر العقد وتموضع BS في النقطة (120,120)

بالضرورة أن يكون مفيداً دائماً نظراً لأنّ بعض التطبيقات تحتاج لأن تكون التغذية الراجعة من العقد موثوقة, ودورة حياة شبكة أكبر ليس بالضرورة أن يكون مفيداً إلّا إذا اقتُرنَ بفترة استقرار أطول.

### 4.1.2. فترة الاستقرار :

وهي الوقت الفاصل بين بدء عمليات الشبكة حتى موت أول عقدة, وتدعى أيضاً منطقة الاستقرار .

نلاحظ من الشكل (6) أنّ فترة الاستقرار في LEACH المحسّن أكبر منها في LEACH في حال كون المحطة الأساسية ضمن الشبكة وبالتالي فإنّ بروتوكول LEACH المُحسّن يعطي أداء أفضل من أجل مقياس فترة الاستقرار .

### 4.1.3. فترة عدم الاستقرار :

تُعرَّف بأنها الفترة مابين موت أول عقدة وآخر عقدة , ويعتبر المقياس ضروري للتطبيقات التي تحتاج إلى وثوقية عالية عند إرسال تقارير (تغذية راجعة) من شبكة الحساسات.

نلاحظ من الشكل (6) أنّ فترة عدم الاستقرار في LEACH المحسّن أصغر منها في LEACH, وبالتالي يُعتبر LEACH المحسّن البروتوكول الأنسب للتطبيقات التي تحتاج لوثوقية.

### 4.1.4. الإنتاجية :

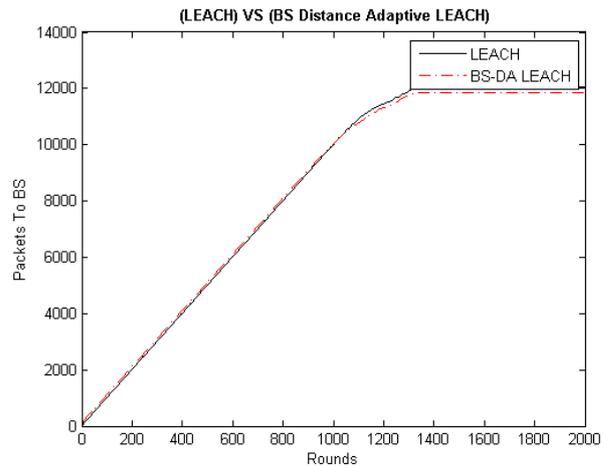

الشكل (7) عدد الطرود المُرسلة إلى المحطة الأساسية

نلاحظ من الشكل (7) أنّ إنتاجيّة LEACH المُحسّن من حيث عدد الطرود المُرسلة إلى المحطة الأساسية أصغر بقليل منها في LEACH (اختلاف بسيط بين انتاجيتي البروتوكولين).

تحسين بروتوكول **LEACH** ليتكيّف مع بعد الحساسات عن المحطة الأساسية

يوضِّح الشكل (10) بأنَّ المحطة الأساسية تتوضَّع خارج حقل, حيث تمّ نشر العقد على مساحة 100*100 وتوضّعت المحطة الأساسية في النقطة (120,120) .

### 4.2.1. دورة حياة الشبكة:

هي الوقت الفاصل بين بدء عمليات الشبكة وحتى موت آخر عقدة حية.

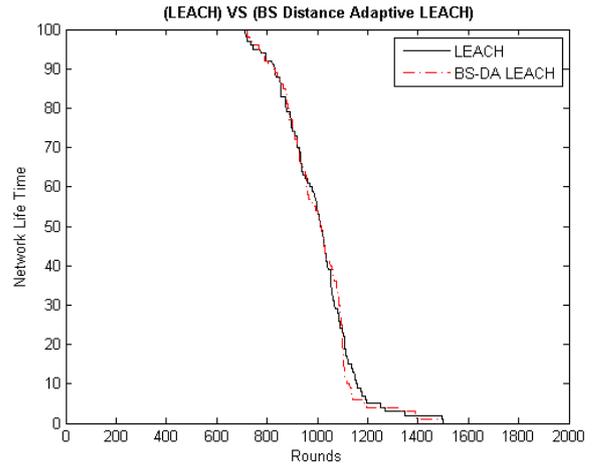

الشكل (11) مقارنة دورة حياة الشبكة بين LEACH و LEACH المحسّن

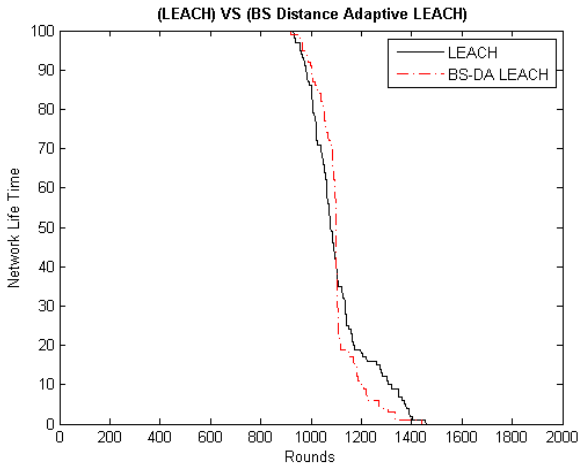

الشكل (13) مقارنة دورة حياة الشبكة بين LEACH و LEACH المحسّن

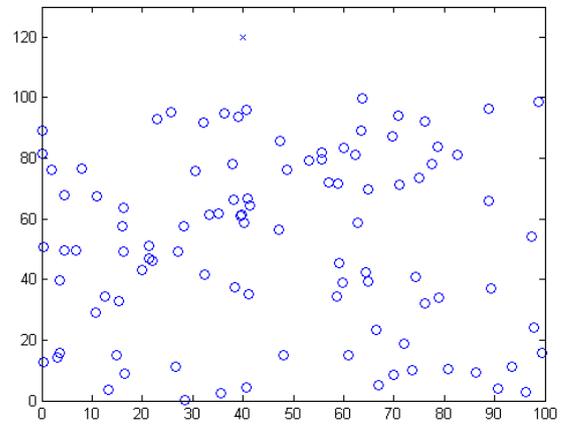

الشكل (12) نشر العقد وتموضع BS في النقطة (40,120)

يوضّح الشكل (12) بأنَّ المحطة الأساسية تتوضَّع خارج حقل, حيث تمّ نشر العقد على مساحة 100*100 وتوضّعت المحطة الأساسية في النقطة (40,120) .

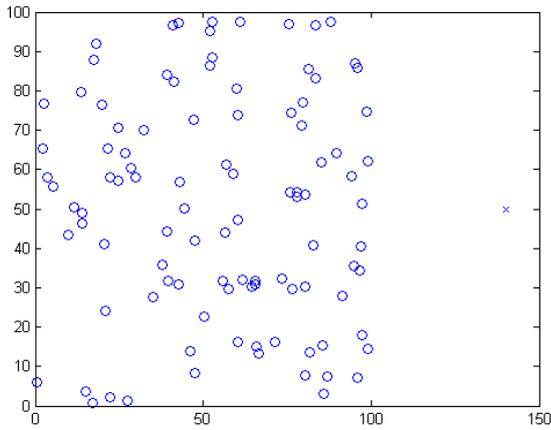

الشكل (14) نشر العقد وتموضع BS في النقطة (140,50)

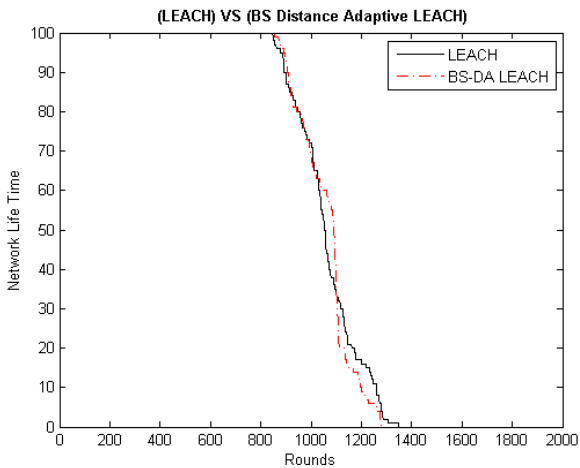

الشكل (15) مقارنة دورة حياة الشبكة بين LEACH و LEACH المحسّن

نلاحظ من الأشكال (11)(13)(15) تطابق دورة حياة الشبكة في كل من LEACH و LEACH المحسَّن مع اختلاف بسيط في بروتوكول LEACH المحسَّن حيث تكون فترة عدم الاستقرار أقصر ويظهر ذلك بشكل واضح في الشكل (15) حيث تكون فترة الاستقرار أطول بعدة جولات.

### 4.2.2. فترة الاستقرار:

نلاحظ تطابق فترة الاستقرار في كل من LEACH و LEACH المحسَّن مع اختلاف بسيط أحياناً حيث يميل LEACH المحسَّن إلى استقرار أكبر كما تُظهر الأشكال (11) و(15).

### 4.2.3. فترة عدم الاستقرار:

كما عرفنا مسبقاً فترة عدم الاستقرار بأنها الفترة مابين موت أول عقدة وآخر عقدة , و هذا المقياس ضروري للتطبيقات التي تحتاج إلى وثوقية عالية عند إرسال تقارير (تغذية راجعة) من شبكة الحساسات, ونلاحظ تطابق فترة عدم الاستقرار في كل من LEACH و LEACH المحسَّن مع اختلاف بسيط في بروتوكول LEACH المحسَّن حيث تكون فترة عدم الاستقرار أقصر.

### 4.2.4.الإنتاجية :

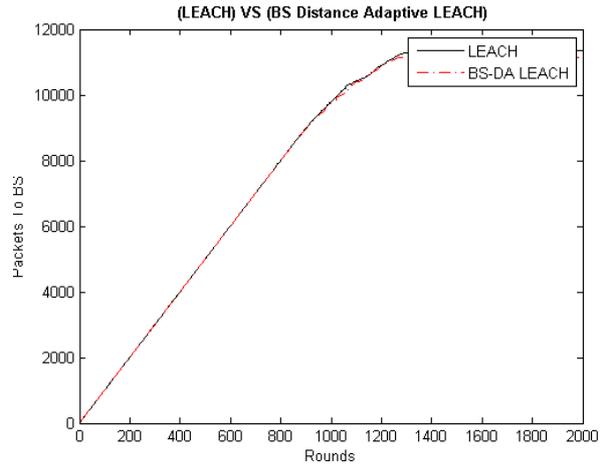

الشكل (16) عدد الطرود المُرسلة إلى المحطة الأساسية

نلاحظ أنّ انتاجية LEACH المُحسّن من حيث عدد الطرود المرسلة إلى المحطة الأساسية تُماثل تقريباً تلك التي في LEACH.

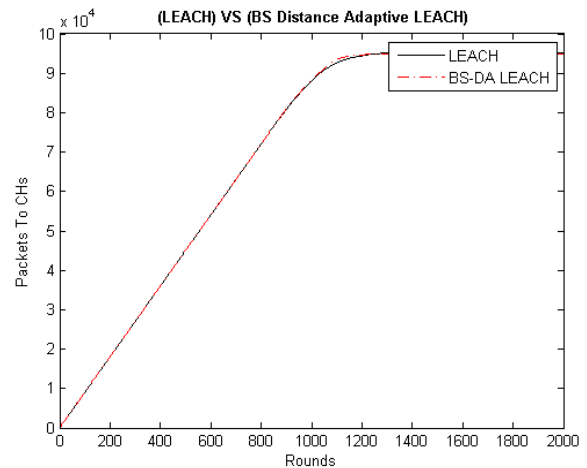

الشكل (17) عدد الطرود المُرسلة إلى رؤوس العناقيد

نلاحظ من الشكل(17) تطابق الإنتاجية من حيث عدد الطرود المُرسلة إلى رؤوس العناقيد في كل من LEACH و LEACH المحسَّن.

### 4.2.5. عدد رؤوس العناقيد في كل جولة:

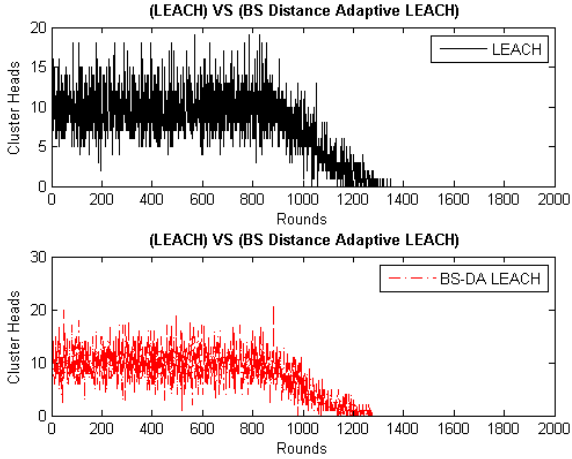

الشكل (18) عدد رؤوس العناقيد بكل جولة

نلاحظ من الشكل(18) تطابق عدد رؤوس العناقيد تقريباً في كل جولة من أجل كلاً من LEACH و LEACH المحسَّن.

## 5. الخلاصة:

قام بروتوكول LEACH المُتكيِّف مع البعد عن المحطة الأساسية (LEACH المُحسّن) بالتحسين على مقياسين مهمين جداً من مقاييس الأداء وهما (فترة الاستقرار وفترة عدم الاستقرار), حيث أعطى هذا البروتوكول فترة استقرار أكبر وفترة عدم استقرار أصغر وذلك في حال توضُّع المحطة الأساسية ضمن حقل العمل, كما خلق توازن في استهلاك الطاقة وتوزيع الإستهلاك بشكل يضمن استقرار الشبكة بشكل أكبر, أما في حالة توضُّع المحطة الأساسية خارج حقل العمل فيعطي أداء مماثل تقريباً لأداء بروتوكول LEACH.

## 6. المراجع:

## 7. مسرد المصطلحات:

| المصلح العربي | المعنى الإنكليزي |
|---|---|
| شبكة الحساسات اللاسلكية | Wireless sensor network |
| المحطة الأساسية أو المصرف | Base Station or Sink node |
| LEACH المحسّن | Base station distance adaptive LEACH |
| فترة عدم الاستقرار | Instability period |
| فترة الاستقرار | Stability period |
| دورة حياة الشبكة | Network lifetime |
| نشر عقد الشبكة | Network deployment |
| النموذج الراديوي | Radio model |
| تغذية راجعة | Feedback |
| معدل طاقة الشبكة المقدّرة | Estimated average energy $\bar{E}(r)$ or $E_a$ |
| العقد الرأسية | Cluster Heads (CH) |
| العقد غير الرأسية | Non-Cluster Head Nodes(Normal Node) |
| التوجيه المسطّح | Flat routing |
| التوجيه الهرمي | Hierarchical routing |
| التوجيه المعتمد على موقع العقد | Location based routing |
| التوجيه المعتمد على جودة الخدمة | QOS based routing |
| برمترات المحاكاة | Simulation parameters |